\documentclass[12pt,preprint]{aastex}

\def\stacksymbols #1#2#3#4{\def\theguybelow{#2}
        \def\verticalposition{\lower#3pt}
        \def\spacingwithinsymbol{\baselineskip0pt\lineskip#4pt}
        \mathrel{\mathpalette\intermediary#1}}
\def\intermediary #1#2{\verticalposition\vbox{\spacingwithinsymbol
        \everycr={}\tabskip0pt
        \halign{$\mathsurround0pt#1\hfil##\hfil$\crcr#2\crcr
                \theguybelow\crcr}}}
\def\lta{\stacksymbols{<}{\sim}{2.5}{.2}}

\shorttitle{Dust in Hot Gas}
\shortauthors{Temi et al.}

\begin{document}

\title{Dust in Hot Gas: Far Infrared Emission from Three 
       Local Elliptical Galaxies}

\author{Pasquale Temi\altaffilmark{1,2}, William G. Mathews\altaffilmark{3}, 
Fabrizio Brighenti\altaffilmark{3,4}, Jesse D. Bregman\altaffilmark{1} }

\altaffiltext{1}{Astrophysics Branch, NASA/Ames Research Center, MS 245-6,
Moffett Field, CA 94035.}
\altaffiltext{2}{SETI Institute, Mountain View, CA 94043.}
\altaffiltext{3}{University of California Observatories/Lick Observatory,
Board of Studies in Astronomy and Astrophysics,
University of California, Santa Cruz, CA 95064.}
\altaffiltext{4}{Dipartimento di Astronomia,
Universit\`a di Bologna, via Ranzani 1, Bologna 40127, Italy.}

\begin{abstract}
We present far-IR ISO observations of three early-type galaxies in the Virgo 
cluster. The data were recorded using the ISOPHOT instrument in both 
the P32 oversampled maps and the P39/37 sparse maps. The maps reach 
the limiting sensitivity of the ISOPHOT instrument at 60, 90,
and 180$\mu$m. Two of the most prominent elliptical galaxies 
in Virgo -- NGC 4472 and NGC 4649 -- clearly 
show no emission at far-IR wavelengths at a level of few tens of mJy,
but NGC 4636 is detected at all three wavelengths.  
We have computed the far-IR emission from dust for NGC4472 and NGC4636
under the assumption that dusty outflows from evolving red giant
stars are continuously supplying dust to the interstellar medium
and that the grains, once diffused into the interstellar medium,
are sputtered away by collisions with ions.
While the calculated fluxes are consistent with the 
observed upper limits for NGC4472, 
the dust emission detected in NGC4636 support 
the hypothesis that additional dust has been accreted in a very recent 
($\lta$ few $10^8$ yrs) merger with a dusty,  
gas-rich galaxy.

\end{abstract}


\section{Introduction}

The interstellar medium (ISM) in elliptical galaxies, although mostly
X-ray emitting hot gas, contains a significant amount of dust and cold
gas of uncertain origin. 
Most of the ISM within the
half-light radius is produced by stellar mass loss.  Since the mean
stellar abundance in typical giant elliptical galaxies is comparable
to solar, envelopes ejected from evolving giant stars should
produce dust just as observed in similar stars in the Milky Way.
In fact, relatively hot circumstellar dust has been observed
in many elliptical galaxies at $\sim 10\mu$m 
with both IRAS and ISO satellites 
\citep{kna89, fer02, ath02}. 
This mid-IR emission has a de Vaucouleurs profile
similar to the background starlight and is therefore 
consistent with circumstellar dust recently 
ejected from evolving red giant stars. 
After about $10^4 - 10^5$ years, the gas and 
dust ejected by giant stars orbiting in elliptical galaxies expands
sufficiently to interact with the ambient hot gas.  According to
currently popular assumptions, as the ejected gas is
dynamically disrupted, it is (shocked and conductively) heated to the
virial temperature ($\sim 10^7$ K) of the hot ISM and the new 
dusty gas merges with its hot environment.  
When surrounded by gas at $\sim 10^7$ K, the dust 
is sputtered until it also merges into the hot gas
typically on a longer timescale.  However, during the sputtering time,
which varies with galactic radius, the dust receives enough energy
from electrons and starlight to emit copiously in the mid and far infrared.\\
A substantial number of elliptical galaxies have been observed using
the ISOPHOT \citep{lem96} instrument on board the ISO
\citep{kes96} satellite.
The spectral region covered by ISOPHOT, between 60 and 200$\mu$m,
together with its mapping capabilities and good angular resolution,
make the instrument well suited for detecting thermal emission from
interstellar dust. In particular the extension of its spectral
coverage to longer wavelengths, relative to the IRAS satellite,
provides a better sampling of the far-infrared emission
where the expected flux density from the cold 
interstellar dust component peaks.
ISOPHOT also has a fainter limiting magnitude compared to IRAS
and diffraction limited imaging that could 
help resolve and disentangle the various expected sources 
of emission: central dust clouds, circumstellar dust,
and truly interstellar dust distributed throughout the
galactic volume.
Here we report on ISOPHOT observations and analyses of three bright
nearby early-type galaxies that are very similar in their optical 
parameters but are significantly different in their far-IR emission.

\section{Observations and Data Reduction}

ISOPHOT observations of early-type galaxies have been taken in
several observing modes (Astronomical Observing Template, referred to
as AOTs), including
the oversampled maps P32, the P37/38/39 sparse maps and P22 multi-filter
photometry data. Here we present data from three very bright
ellipticals located in the Virgo cluster: NGC4636, NGC4649 and NGC4472. \\
NGC4636 has been observed using the C100 3$\times$3 and the C200
2$\times$2 detector arrays with the sparse map on-off mode (AOT P37/39)
at 60 and 100$\mu$m (C100) and at 180$\mu$m (C200).
Observations were made in each filter with one on-source single point staring
mode and one off-source exposure. 
The background position was located  $\sim12^\prime$
north of the target, on a blank sky position. Data reduction and calibration
were carried out using the PHT Interactive Analysis tool PIA, version 9.1
\citep{gab97} together with the calibration data set V4.0
\citep{lau98}. The reduction
included correction for the non-linear response of the detectors, readout
deglitching, and linear 
fitting of the signal ramps. After resetting of all ramp
slopes and subtracting the dark current, the calibration of detector
responsivity and its changes was performed using measurements of the
thermal fine calibration source (FCS) on board.
The sky subtracted on source signal 
was then corrected for the fraction of the point-spread
function included within the field of view of the detectors. 
Finally, the flux densities were extracted 
applying a color correction assuming that the SED of
the galaxy can be approximated with a blackbody of temperature 30 K.
The statistical
errors derived from signal processing are about 5-20\%, depending on
the wavelength range and object brightness, but systematic errors due
to absolute calibration accuracy are estimated to be 30\% \citep{kla98}.
The derived flux density for NGC4636 in the three filters is shown
in Table~\ref{tbl-1}. Values are in good agreement with previous IRAS observations.\\
NGC4649 and NGC4472 have been observed in the P32 mode in three broadband filters
at 60, 90 and 180$\mu$m using the C100 and C200 detectors. The maps were obtained
scanning the spacecraft Y and Z axes in a grid of 4$\times$4 (NGC4649) and
5$\times$4 (NGC4472) points. The focal plane
chopper was stepped at intervals of one-third of the detector pixel at each
raster position, providing a sky sampling in the Y direction of 15$^{\prime\prime}$
and 30$^{\prime\prime}$
for the C100 and C200 detectors. All the maps were centered
on the position of the nuclei of the two galaxies covering a field of
about 9$^\prime\times 6^\prime$. To reduce the P32 data we used a dedicated
software package (P32TOOLS) developed at MPI Kernphysik in Heidelberg, and supported by
the VILSPA ISO Data Center \citep{tuf03}. The new routines
allow a proper correction for transients in PHT32 measurements. The ISOPHOT
C200 and especially the C100 detectors have a complex non-linear response as a function
of illumination history on time scales of  0.1-100 sec, depending on the
absolute flux level as well as the flux changes involved. Since no theory
exists to predict the detector response behavior, an empirical model has been
found to apply the transient correction  to the data. The severity of the problem
depends on both the source/background ratio and the dwell time at each pointing
direction. Generally, ISOPHOT data are reduced in one pass from the raw input data
with full time resolution to the final calibrated map. To correct for responsivity
drift effects, however, it is necessary to iterate between a sky map and the
input data at full time resolution. This different concept
has been implemented in the dedicated data reduction package.
The implementation of these new algorithms has been already
successfully applied to an extended sample of Virgo cluster galaxies by
\citet{pop02}. 
Great effort has been devoted to the removal of spikes and
longer lived tails in the detector responsivity that are often generated by
cosmic rays hitting the detector pixels. For the observations presented here,
which are faint discrete sources, an accurate deglitching procedure
becomes critical in determining the instrumental sensitivity.
Although P32TOOLS applies specific deglitching procedures to
remove glitches from the data, some residuals are still present in the maps
made with the C100 detector. Removal of these artifacts was fully achieved for
the C200 detector.\\
The photometric calibrations of discrete sources performed with the ISO flux
scale have been tested against integrated flux densities measured by IRAS at 60
and 100$\mu$m bands on  Virgo cluster galaxies \citep{tuf02}. A very good
linear correlation is found between the ISO and IRAS fluxes [see Figures 7{\it a}
and 7{\it b} in \citet{tuf02}].\\
Elliptical galaxies NGC4472 and NGC4649 were not 
detected at any of the three wavelengths observed by ISOPHOT (60, 90,
and 180$\mu$m).
Figure 1 shows the 90$\mu$m maps for these two galaxies;
the bright source
at the edge  of the NGC4649 map is its spiral companion NGC4647 and it is
prominent at all three wavelengths. The angular resolution of the
oversampled  maps clearly resolves NGC4647 showing that there is no
contribution to the signal associated with 
NGC4649. 
In the literature, IRAS detections at 60 and 100$\mu$m have been  
erroneously attributed to NGC4649, simply because the spatial
resolution of the IRAS observations, 3\arcmin - 5\arcmin ~at 100$\mu$m,
is very large and comparable to the optical size of most elliptical galaxies.
The maps reach the limiting sensitivity of the ISOPHOT instrument at a
level of a few tens of mJy and the 3$\sigma$ upper limits are presented
in  Table~\ref{tbl-1}.

\section{A MODEL FOR NGC 4472 AND NGC 4636}
We estimate the FIR emission from dust dispersed in the ISM
along the lines of Tsai \& Mathews (1995, 1996). A full description 
of the method will be presented elsewhere \citep{tem03}
In brief, we assume that the grains
are ejected from evolving stars and mix with the hot ISM in a short
time, $\sim 10^5-10^6$ yr \citep{mat90}. 
Once diffused into the hot gas, the grains 
are sputtered at a rate
\begin{equation}
{da \over dt} = - 3.2 \times 10^{-14} n_p [ 1 +
(2 \times 10^6 / T)^{2.5} ]^{-1}
~~\mu{\rm m}~{\rm s}^{-1},
\end{equation}
where $a$ is the grain radius (in $\mu$m) and $n_p$ is the proton density
\citep{dra79, tsa95}

Let $N(r,a)da$ be the steady state 
number of grains per cm$^3$ with radius between
$a$ and $a+da$ at galactic radius $r$. 
\citet{tsa95} showed that the grain population 
must satisfy 
\begin{equation}
{\partial \over \partial a} \left[ N \left( {da \over dt}
\right) \right] = S(r,a),
\end{equation}
where $S(r,a)da$ is the rate  
at which grains with radii $a$ are expelled from stars per cm$^3$.
Mid-IR (6 - 15$\mu$m) ISO observations of a sample of 
elliptical galaxies, including NGC4472, detect strong
dust emission that can be understood as emission
from circumstellar dust from mass-losing AGB stars
at or near their
globally expected rate \citep{ath02}. We assume
$S(r,a)=S_o(r) a^{-s}$ for $0<a<a_{\rm max}$, with $s=3.5$ 
\citep{mat77}, so the coefficient 
\begin{equation}
S_o(r) = {3 \delta \alpha_* \rho_* \over
4 \pi \rho_g 10^{-12}} { (4 - s) \over a_{max}^{4-s}}
\end{equation}
depends on the rate that the old stellar
population ejects mass
$\alpha_* = 4.7 \times 10^{-20}$ s$^{-1}$, 
the stellar density 
$\rho_*(r)$, the initial dust to gas 
mass ratio scaled
to the stellar metal abundance
$\delta = (1/150)z(r)$,
and the density of silicate grains,
$\rho_g = 3.3$ gm cm$^{-3}$. The solution of Equation (2) is then
\begin{equation}
N(r,a) = \left| {da \over dt} \right|^{-1}
{S_o(r) \over (s - 1) } a^{1-s}~~~a \leq a_{max}.
\end{equation}

The grain temperature is determined by the balance of heating (by both
absorption of stellar radiation and by electron-grain collisions)
and cooling:
\begin{equation}
\int_0^{\infty} 4 \pi J_*(r,\lambda) Q_{abs}(a,\lambda)
\pi a^2 d \lambda
+ 4 \pi a^2 (1/4) n_e \langle v_e E_e \rangle
\tau(a)
= 4 \pi a^2 \sigma_{SB} T_d(r,a)^4
\langle Q_{abs} \rangle (T_d, a). 
\end{equation}
Since grains experience a relatively high collision frequency 
with electrons and photons, 
we have not considered stochastic temperature excursions of 
individual grains \citep{tsa95}.
Since the 9.7$\mu$m silicate feature
is seen in emission in 
a number of elliptical galaxies, including
NGC4472 \citep{ath02}, we
assume that all grains have properties similar to astronomical
silicates \citep{lao93}.
Moreover, to simplify the calculations we approximate 
the absorption coefficient with $Q_{abs} \approx a \psi(\lambda)$ 
and the Planck-averaged values with 
$\langle Q_{abs} \rangle \approx 1.35 \times 10^{-5} T_d^2 a$
[see \citet{dra84}].
The mean intensity of starlight
is found by integrating over a de Vaucouleurs
stellar profile modified with a core of slope $\rho_* \propto
r^{-p}$ within $r_c$.
At a distance $d = 17$ Mpc, NGC4472 has an effective radius
$R_e = 8.57$ kpc, total stellar mass $M_{*,t} = 7.26 \times
10^{11}$ $M_{\odot}$ [assuming
$M_{*,t}/L_B = 9.2$ \citep{van91}], core radius $r_c=200$ pc and
$p=0.9$ \citep{fab97}. For NGC 4636, assuming again
$d = 17$ Mpc, $R_e = 8.32$ kpc, $M_{*,t} = 3.80 \times 10^{11}$ $M_{\odot}$,
$M_{*,t}/L_B = 10.74$, $r_c=260$ pc, $p=1.0$.
The collisional heating term in Equation (5) includes a correction
$\tau(a)$ for small grains which do not completely stop the
electrons \citep{dwe86} and $\langle v_e E_e \rangle = 
\sqrt{32/\pi m_e} (k T)^{3/2}$ is an average over a Maxwellian distribution.
The ISM electron density and temperature are
calculated using the fitting formulae given in \citet{bri97}.

Finally the FIR emissivity is calculated from 
\begin{equation}
j(r,\lambda) = {1 \over 4 \pi} \int_{0}^{a_{max}}
N(r,a) 4 \pi a^2 10^{-8} Q_{abs}(a,\lambda) \nonumber
\times \pi B(T_d(r,a),\lambda) da.
\end{equation}

Figure 2 shows the resulting dust temperature profile for NGC 4636,
for grains of three different sizes. Slightly higher temperatures
are predicted for NGC 4472. We find that collisional heating
dominates over stellar radiation heating for grain sizes $a\leq 0.3$
$\mu$m and $a\leq 1$ $\mu$m for NGC 4472 and NGC 4636 respectively.

The computed far-IR spectra calculated for $a_{\rm max}=1$ $\mu$m
and $a_{\rm max}=0.1$ $\mu$m are shown in Figure 3,
together with the ISO detections or upper limits.
For NGC 4472 the predicted fluxes are consistent with 
the ISO upper limits provided
$a_{\rm max}\leq 1$ $\mu$m. Thus, no strong constraints on the properties
of interstellar dust can be derived for this galaxy.
On the other hand, for NGC4636 the detected fluxes
are a factor $\sim 50$ above the computed values
at all three observed wavelengths, assuming $a_{\rm max}=1$ $\mu$m.
The shape of the calculated spectral energy distribution,
however, fits the observations quite well,
indicating that the excess emission is
caused by a larger amount of dust rather than by hotter grains.
The calculated flux ratios
$F_{90\mu{\rm m}}/F_{60\mu{\rm m}} = 2.7$ and $3.8$
($a_{\rm max}=0.1$ and $1$ $\mu$m);
$F_{180\mu{\rm m}}/F_{90\mu{\rm m}} = 1.2$
and $2.1$ ($a_{\rm max}=0.1$ and $1$ $\mu$m)
compare reasonably well
with the observed values 
$F_{90\mu{\rm m}}/F_{60\mu{\rm m}}\sim 2.6$
and $F_{180\mu{\rm m}}/F_{90\mu{\rm m}}\sim 1.6$.
The model for NGC 4636 predicts a total mass of interstellar dust
$M_{\rm dust}=
1.7 \times 10^5$ M$_\odot$ if $a_{\rm max}=1$ $\mu$m or
$M_{\rm dust}=1.7 \times 10^4$ M$_\odot$ for $a_{\rm max}=0.1$ $\mu$m.
We are led to speculate that NGC 4636 has accreted dust
in a recent merger with a gas-rich galaxy.
Other explanations appear less palatable. A higher (by a factor $\geq 50$) 
dust production rate
by a younger stellar population would be inconsistent with the
optical colors of NGC4636. Moreover, we found that
the underluminosity of
internally produced
dust with respect to the observed values cannot be eliminated by varying
the grain size distribution parameters
$a_{\rm max}$ and $s$.

\section{Conclusions}
We have presented ISO observations of three bright elliptical galaxies
in the Virgo cluster. While NGC 4472 and NGC 4649 emit no
far-IR emission at the limiting sensitivity of the instrument, NGC 4636
is clearly detected.
These observations have been interpreted with theoretical models 
for the far-IR emission from interstellar dust under the assumption
that the circumstellar dust observed
in the mid-IR \citep{ath02}
ultimately becomes interstellar and is
exposed to the mean galactic starlight and
to bombardment by thermal electrons in the hot
gas. Grains receive somewhat more
energy from electrons than from stellar photons.
The calculated far-IR spectrum peaks at
$\lambda \sim 180$ $\mu$m, a spectral region not accessible to IRAS,
but well covered by ISO. \\
Although NGC 4472 and NGC 4636 are similar
in terms of mass, luminosity, hot gas content, 
and other physical parameters, they show a remarkable discrepancy
in their far-IR emission. While the upper limits for NGC 4472
are consistent with emission from dust produced by the evolving
stellar population, the model for NGC 4636 falls short by $\sim 50$
in explaining the observed fluxes. If the dust mixture would include
a significant amount of graphite grains, the FIR emission would
be at most a factor of few larger (Tsai \& Mathews 1996), still
much too low to explain the observations. 
This strongly suggests that dust has been accreted
by NGC4636 in a recent merging event with a gas-rich galaxy.
A dust mass larger by a factor of about 50-70 is needed in order to match
the observed far-IR luminosity, if its relative spatial distribution is
similar to the one calculated.
If the required extra-dust has an external
origin, however, its distribution may be
irregular and patchy. Without spatial information on the FIR
emission we are not in the position to constrain the spatial
location of the external dust.
\citet{rav01} find no compelling evidence for dust extinction in 
NGC4636 but the morphological 
properties of its stellar core are deviant.
An asymmetrical distribution of 
diffuse H$\alpha$ + [NII] emission 
with an irregular velocity field has been detected by \citet{cao01}
in the central region of NGC 4636.
The very disturbed hot ISM \citep{jon02} 
also suggests a recent merging event 
and this is further supported by the apparent failure 
of the hot gas within $\sim R_e/2 = 4$ kpc to be in hydrostatic 
equilibrium \citep{bri97}.\\
The hypothetical merger must be very recent because 
of the short sputtering time. 
>From Equation (1) 
the sputtering time for grains of radius $a_{\mu}$ (in m$\mu$) is 
$t_{sputt} \approx a_{\mu}/|d a_{\mu}/dt |
\approx 1.2 \times 10^6 a_{\mu} n_e^{-1}$ yrs.
Since the gas density in NGC4636 varies as
$n_e \approx 0.178 r_{kpc}^{-1.36}$ cm$^{-3}$ \citep{bri97},
we find  
$t_{sputt} \approx 1.2 \times 10^8 a_{\mu} (r_{kpc}/R_{e,kpc})^{1.36}$
yrs.
The total mass of dust that merged is $\sim 10^6 - 10^7$ $M_{\odot}$ 
and the associated gas would have mass $\sim 10^8 - 10^9$ 
$M_{\odot}$.
For this gas to have been heated to $\sim 10^7$ K 
and for some of the dust to have survived, 
the dusty merger must have 
occurred less than $t_{sputt}$, or a few $10^8$ yrs ago. 
The total mass of hot gas in NGC4636 
within $\sim R_e/2$ where hydrostatic 
equilibrium fails, $\sim 2 \times 10^8$ $M_{\odot}$, 
is comparable to that of the added gas, 
so considerable dynamical disruption is possible. 
Hydrostatic equilibrium would be reestablished in 
a few dynamical times which is  
$\sim 3 \times 10^7$ yrs at $r \sim R_e/2 = 4$ kpc.
Dust survival and continued dynamical activity both
indicate that an important merger
may have occurred within the last few $10^8$ yrs. 
It would be interesting to seek evidence for 
the merged stars in the stellar spectrum of NGC4636.

\acknowledgments
This study is supported by NASA grant 399-20-01 
for which we are very grateful.
WGM is supported by grants from the NSF and NASA.

\clearpage

\figcaption[fig2a.ps,fig2b.ps]{Optical  image (contours) superimposed to 
ISOPHOT 90$\mu$m map for NGC4649 (left panel) and NGC4472 (right panel). 
Both galaxies have not been detected at 60, 90, and 100$\mu$m. The bright source
at the edge  of NGC4649 map is its spiral companion NGC4647. \label{fig1}}

\figcaption[fig3.eps]{Dust temperature for NGC4636 due to starlight and
electron-grain collisions computed as a function of radial distance.
For grains with size $a\leq 0.3$ $\mu$m the collisional
heating is larger than radiative heating.   \label{fig2}}

\figcaption[fig4.eps]{Computed Far-IR flux from NGC4636
and NGC4472  calculated for $a_{max} = 1\mu$m (solid lines) and
$a_{max} = 0.1\mu$m (dashed lines).
The ISO detections are shown as open triangles for NGC4472 (upper limits),
and open circles for NGC4636. \label{fig3}}

\newpage
\plotone{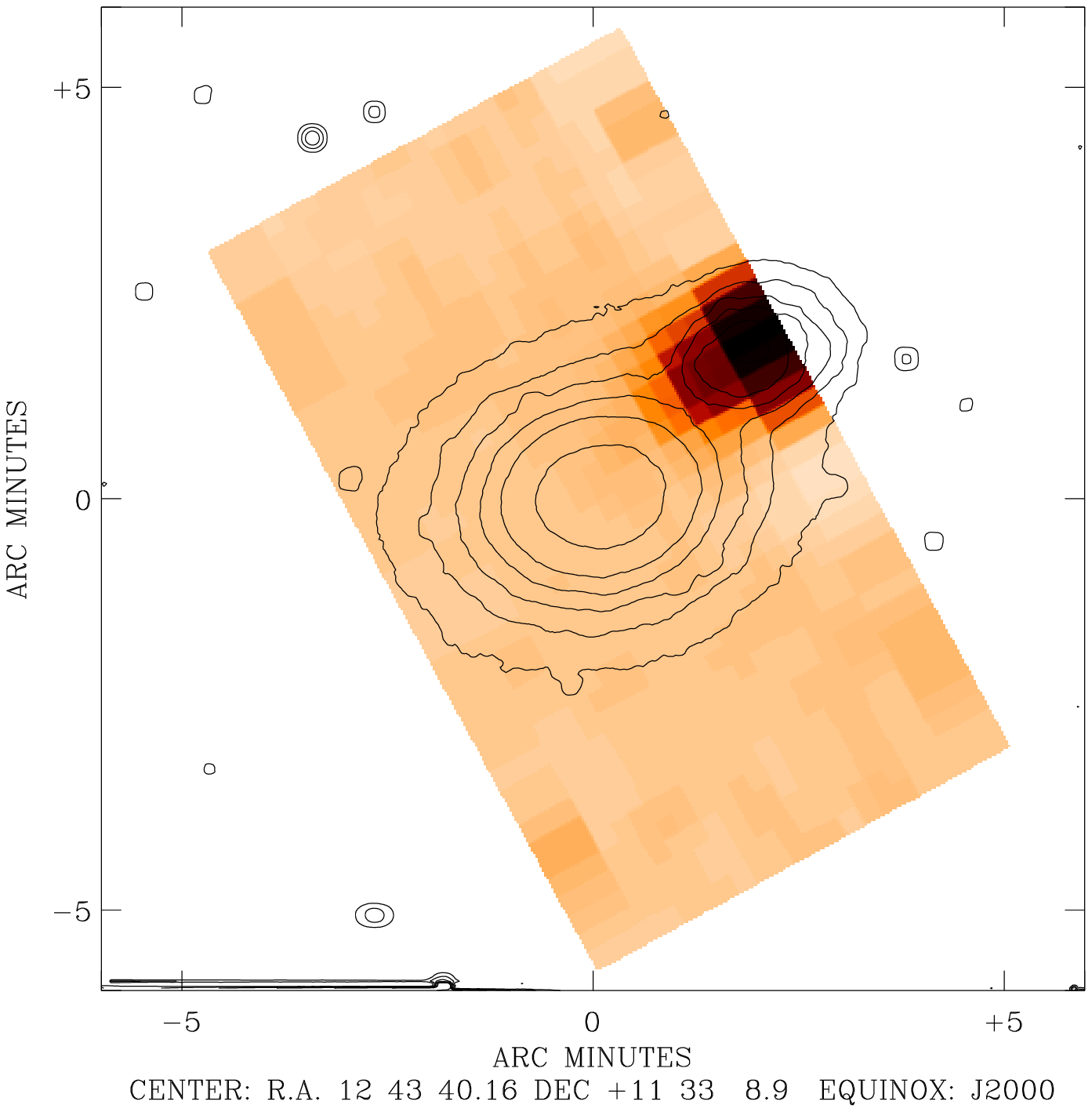}
\newpage
\plotone{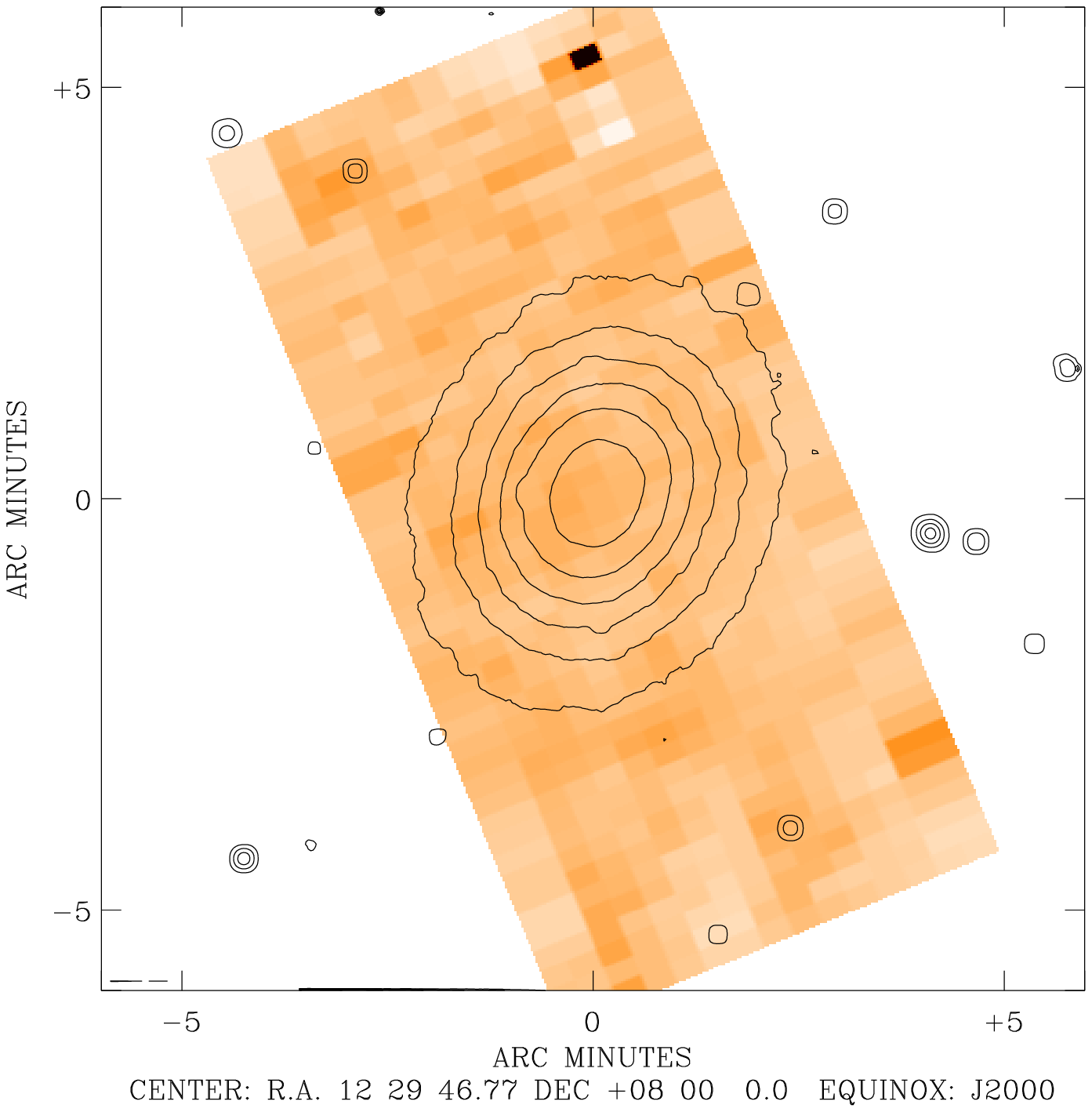}
\newpage
\plotone{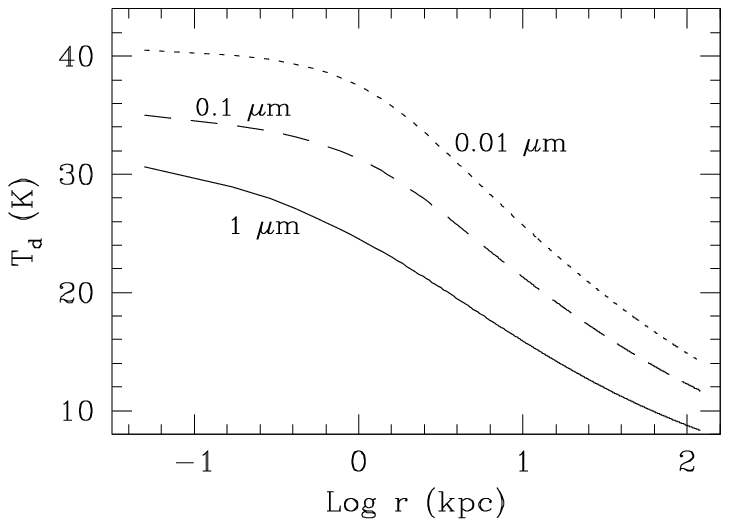}
\newpage
\plotone{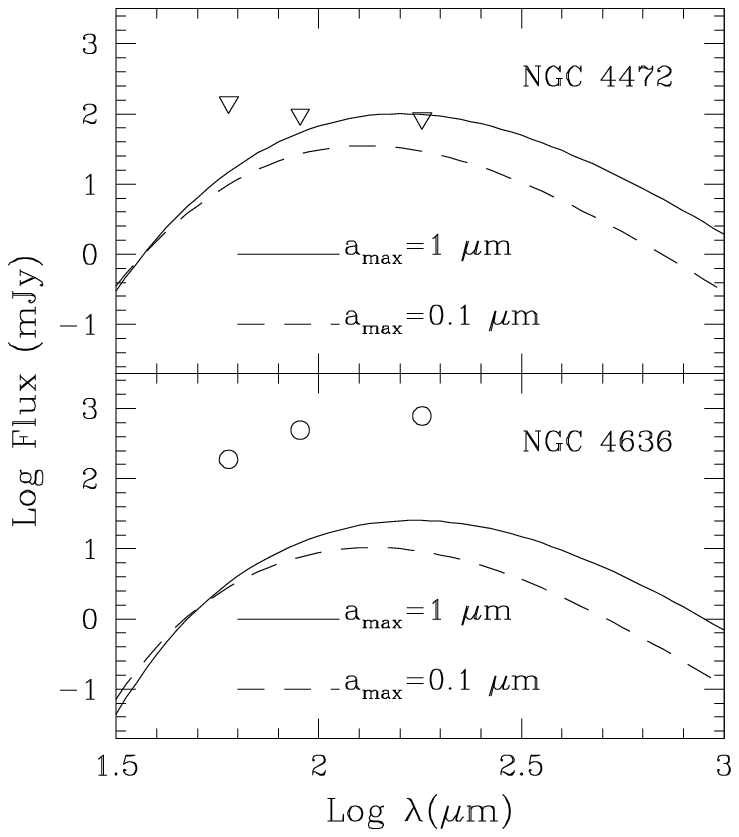}

\clearpage

\begin{deluxetable}{cccrcc}
\tablewidth{0pt}
\tablecaption{ISOPHOT photometry. \label{tbl-1}}
\tablecolumns{6}
\tablehead{
\colhead{Galaxy}            & \colhead{log (L$_{\rm B}$)} &
\colhead{log (L$_{\rm x}$)} & \colhead{${\rm \lambda}$}   &
\colhead{Flux (ISO)}        & \colhead{Flux (IRAS)}   \\ 
\colhead{}                  & \colhead{(L$_{B \odot}$)}   &
\colhead{(erg s$^{-1}$)}    & \colhead{($\mu$m)}          &
\colhead{(mJy)}             & \colhead{(mJy)}
}            
\startdata
NGC 4649\tablenotemark{a}&10.73 & 41.21 & 60 & $\leq$137\tablenotemark{c} & 900 $\pm$ 65\tablenotemark{d} \\
                         &      &       & 90 & $\leq $ 85\tablenotemark{c}& \nodata  \\
                         &      &       &100 &   \nodata                  &2310 $\pm$ 270\tablenotemark{d} \\
                         &      &       &180 & $\leq$110\tablenotemark{c} & \nodata   \\
NGC 4472\tablenotemark{a}&10.90 & 41.66 & 60 & $\leq$147\tablenotemark{c} &   0 $\pm$ 66 \\
                         &      &       & 90 & $\leq$ 99\tablenotemark{c} & \nodata  \\
                         &      &       &100 &         \nodata            &   0 $\pm$  94 \\
                         &      &       &180 & $\leq$ 87\tablenotemark{c} & \nodata   \\
NGC 4636\tablenotemark{b}&10.55 & 41.58 & 60 & 187 $\pm$ 57               & 200 $\pm$ 34 \\
                         &      &       & 90 &   491 $\pm$ 64             & \nodata  \\
                         &      &       &100 &   \nodata                  & 560 $\pm$ 179 \\
                         &      &       &180 &  790 $\pm$ 71              & \nodata   \\
 \enddata
\tablenotetext{a}{Observed in the P32 oversampled map mode}
\tablenotetext{b}{Observed in P37/39  one position sparse map mode}
\tablenotetext{c}{No detection, 3 $\sigma$ upper limit flux}
\tablenotetext{d}{The aperture photometry includes the spiral companion NGC4647}
\end{deluxetable}

\end{document}